\documentclass[superscriptaddress,pra,twocolumn,amsmath,amssymb]{revtex4}
\usepackage{mathrsfs}
\usepackage{amssymb, amsbsy, amsmath, latexsym, dsfont, array, layout, graphicx,mathrsfs,color,soul,hyperref}

\newcommand{\one}{\mathds{1}}
\newcommand{\ket}[1]{\left|{#1}\right\rangle}
\newcommand{\bra}[1]{\left\langle{#1}\right|}

\definecolor{golden}{rgb}{0.8,0.6,0.1}

\newcommand{\erf}[1]{Eq.~(\ref{#1})}

\newcommand{\innertime}[2]{\langle #1|#2\rangle}

\begin{document}
\title{Experimental generalized contextuality with single-photon qubits}

\author{Xiang Zhan}
\affiliation{Department of Physics, Southeast University, Nanjing 211189, China}

\author{Eric G. Cavalcanti}
\affiliation{Centre for Quantum Dynamics, Griffith University, Gold Coast, QLD 4222, Australia}

\author{Jian Li}
\affiliation{Department of Physics, Southeast University, Nanjing 211189, China}

\author{Zhihao Bian}
\affiliation{Department of Physics, Southeast University, Nanjing 211189, China}

\author{Yongsheng Zhang}
\email{yshzhang@ustc.edu.cn}
\affiliation{Key Laboratory of Quantum Information, University of Science and Technology of China, CAS, Hefei 230026, China}
\affiliation{Synergetic Innovation Center of Quantum Information and Quantum Physics, University of Science and Technology of China, Hefei 230026, China}

\author{Howard M. Wiseman}
\email{h.wiseman@griffith.edu.au}
\affiliation{Centre for Quantum Dynamics, Griffith University, Gold Coast, QLD 4222, Australia}
\affiliation{Centre for Quantum Computation and Communication Technology (Australian Research Council), Griffith University, Brisbane 4111, Australia}

\author{Peng Xue}
\email{gnep.eux@gmail.com}
\affiliation{Department of Physics, Southeast University, Nanjing 211189, China}
\affiliation{State Key Laboratory of Precision Spectroscopy, East China Normal University, Shanghai 200062, China}

\begin{abstract}
Contextuality is a phenomenon at the heart of the quantum mechanical departure from classical behaviour, and has been recently identified as a resource in quantum computation. Experimental demonstration of contextuality is thus an important goal. The traditional form of contextuality -- as violation of a Kochen-Specker inequality -- requires a quantum system with at least three levels, and the status of the assumption of determinism used in deriving those inequalities has been controversial. By considering `unsharp' observables, Liang, Spekkens and Wiseman (LSW) derived an inequality for generalized noncontextual models that doesn't assume determinism, and applies already to a qubit. We experimentally implement the LSW test using the polarization states of a heralded single photon and three unsharp binary measurements. We violate the LSW inequality by more than 16 standard deviations, thus showing that our results cannot be reproduced by a noncontextual subset of quantum theory.
\end{abstract}

\maketitle

\section{Introduction}

There are a number of proposals for tests which pit quantum mechanics against alternative views of reality, including the theorems of Bell~\cite{Bell} and of Kochen and Specker (KS)~\cite{KS}. Corresponding experimental tests~\cite{exp1,exp2,K+09,ARBC09,B+09,MRCL10} have been performed and support the validity  of quantum mechanics. Bell's theorem refers to a situation with two or more spatially separate particles and states that local hidden variable theories are incompatible with the statistical predictions of quantum mechanics.
The KS theorem has the advantage of applying to a single system, and states that noncontextual hidden variable theories are incompatible with quantum predictions, under the assumption that the measurements can be described by projectors.
A qutrit (three-level system) and five projectors are required for a proof of the traditional KS contextuality in a state-dependent manner~\cite{L+11,KCBS08}, while a qutrit and thirteen projectors for such a proof in a state-independent manner~\cite{P91,B96,BBCP09,YO12,CAB+12,ZWD+12,ZUZ+13}.

To find simpler proofs of contextuality, applicable to a qubit (two-level system), generalizations of KS noncontextuality
have been proposed~\cite{B03,C03,Ara03,Me07}. These all utilise generalized measurements, described by positive operator-valued measures (POVMs). It has been argued, however~\cite{Spe13} that these works make an unwarranted assumption of determinism for unsharp measurements.

More recently, Liang, Spekkens and Wiseman (LSW)~\cite{LSW} (Sec.~7.3) followed a different approach to derive noncontextuality inequalities for a particular class of non-projective measurements.  The relevant class is the unsharp projective measurements, in which each of the set of orthogonal projectors is mixed in some ratio with other projectors from the same set, in order to make the POVM. (Thus each element of the POVM commutes with each other element, just as for a projective measurement.) The LSW assumption is that the response function is likewise a mixture of the deterministic response functions assumed by KS for projective measurements, in the same ratios. Using this principle, LSW derived a generalized noncontextuality inequality involving three different unsharp projective measurements on a qubit. Subsequently, Kunjwal and Ghosh~\cite{KG14} found a triple of unsharp observables that, according to the predictions of quantum mechanics, would give a significant violation of the LSW inequality, in a state-dependent manner.

Here, we experimentally violate the LSW inequality for the first time, via three unsharp binary qubit measurements that are pairwise jointly measurable. We use a photon polarization qubit, and the scheme of Ref.~\cite{KG14}. Our work verifies experimentally that even a single qubit is enough to demonstrate quantum contextuality, under the weak assumptions of Ref.~\cite{LSW}. As we assume the validity of operational quantum theory for the error analysis, our work demonstrates that our results cannot be reproduced by a noncontextual fragment of quantum theory -- an important experimental benchmark.  We exceed the LSW bound by many standard deviations, in an experimentally verified regime of validity for the inequality.

We note that an independent experimental demonstration of contextuality with qubit systems, following techniques complementary to the present work, is reported in~\cite{Mazurek:2016aa}. There, the state preparations and measurements are realized with time-sharing methods, and the problem of noises in measurements is solved with a technique derived within the framework of generalised probabilistic theories.

\section{Theoretical Idea}

\subsection{Scheme for violating the LSW inequality}
A generalized noncontextual model, referred to as a LSW model, can be realized using noisy spin-$\frac{1}{2}$ observables~\cite{LSW}. Specifically, three
such observables, ${\cal M}_k$ ($k=1,2,3$), are required, each described by a two-outcome POVM, ${\cal M}_k=\{E_+^k,E_-^k\}$, of the form~\cite{LSW}
\begin{equation} \label{one-tr}
E_{\pm}^k\equiv \frac{1}{2}\one\pm \frac{\eta}{2} \vec{\sigma}\cdot \hat{n}_k=\frac{1-\eta}{2}\one+\eta \Pi_{\pm}^k.
\end{equation}
Here $\one$ is the $2 \times 2$ identity matrix, $\vec{\sigma}$ is the vector of Pauli matrices $(\sigma_x,\sigma_y,\sigma_z)$, $\hat{n}_k$ is the axis for measurement $k$, and
$\eta\in \left[0,1\right]$ is the sharpness associated with each observable.
For $\eta = 1$, these reduce to projective measurements, ${\cal P}_k=\{\Pi_+^k,\Pi_-^k\}$.
In our experiment, we choose a special case of trine spin axes
\begin{equation} \label{trine}
\hat{n}_1=(0,0,1),\hat{n}_2=(\frac{\sqrt{3}}{2},0,-\frac{1}{2}),\hat{n}_3=(-\frac{\sqrt{3}}{2},0,-\frac{1}{2}),
\end{equation}
equally spaced in the $z$-$x$ plane.

Testing the LSW inequality for a quantum mechanical violation requires a special kind of joint measurability, denoted by joint measurability  contexts $\{\{{\cal M}_1,{\cal M}_2\},\{{\cal M}_2,{\cal M}_3\},\{{\cal M}_1,{\cal M}_3\}\}$. That is,
the three observables ${\cal M}_k$ ($k=1,2,3$)
are pairwise jointly measurable, for all three pairs, but not triply
jointly measurable. Pairwise joint measurability is possible only if $\eta\leq(\sqrt{3}-1) \approx 0.732$~\cite{LSW}.
Triple-wise joint measurability ---
which would eliminate any possibility of contextuality since
the entire experiment could be performed
using a single context $\{{\cal M}_1,{\cal M}_2,{\cal M}_3\}$
--- is possibly only if $\eta < 2/3$~\cite{LSW}. Here we restrict our consideration of $\eta$ to
the narrow range $2/3<\eta\leq(\sqrt{3}-1)$.

The joint measurability context $\{{\cal M}_i, {\cal M}_j\}$ means that there exists a POVM  ${\cal J}_{ij} \equiv\{G_{++}^{ij},G_{+-}^{ij},G_{-+}^{ij},G_{--}^{ij}\}$ satisfying the marginal condition that $\sum_{\varepsilon} G_{\epsilon\varepsilon}^{ij}=E_{\epsilon}^{i}$, and $\sum_{\epsilon} G_{\epsilon\varepsilon}^{ij}=E_{\varepsilon}^{j}$,
where $\epsilon,\varepsilon\in\{+1,-1\}$.
We follow Ref.~\cite{KG14} in using joint POVMs with
 the following general form:
\begin{align}
&G_{++}^{ij}=\frac{1}{2}\left\{\frac{\alpha_{ij}}{2}\one+\vec{\sigma}\cdot\frac{1}{2}\left[\eta\left(\hat{n}_i+\hat{n}_j\right)-\vec{a}_{ij}\right]\right\},\nonumber\\
&G_{+-}^{ij}=\frac{1}{2}\left\{\left(1-\frac{\alpha_{ij}}{2}\right)\one+\vec{\sigma}\cdot\frac{1}{2}\left[\eta\left(\hat{n}_i-\hat{n}_j\right)+\vec{a}_{ij}\right]\right\},\nonumber\\
&G_{-+}^{ij}=\frac{1}{2}\left\{\left(1-\frac{\alpha_{ij}}{2}\right)\one+\vec{\sigma}\cdot\frac{1}{2}\left[\eta\left(-\hat{n}_i+\hat{n}_j\right)+\vec{a}_{ij}\right]\right\},\nonumber\\
&G_{--}^{ij}=\frac{1}{2}\left\{\frac{\alpha_{ij}}{2}\one+\vec{\sigma}\cdot\frac{1}{2}\left[\eta\left(-\hat{n}_i-\hat{n}_j\right)-\vec{a}_{ij}\right]\right\}, \label{Gij}
\end{align}
where $\alpha_{ij}\in \mathbb{R}$ and $\vec{a}_{ij}\in \mathbb{R}^3$, and the relation $G^{ij}_{\epsilon\varepsilon}=G^{ji}_{\epsilon\varepsilon}$ with $\epsilon, \varepsilon\in\{+1,-1\}$ is satisfied.

The LSW inequality is the following~\cite{LSW}
\begin{equation}
R_3\equiv\frac{1}{3}\sum_{(ij)\in\left\{(12),(23),(13)\right\}}\text{Pr}(X_i\neq X_j|{\cal J}_{ij})\leq1-\frac{\eta}{3},
\end{equation}
where $\text{Pr}(X_i\neq X_j|{\cal J}_{ij})$ denotes the probability of obtaining anticorrelated outcomes in a joint measurement denoted ${\cal J}_{ij}$. Note that by the (unreasonable) assumption of outcome determinism for POVMs in Refs.~\cite{B03,C03,Ara03,Me07},
the bound on the right-hand-side would be 2/3~\cite{LSW}, whereas
the LSW bound is {\em at least} $0.756$ (since we require $\eta< 0.732$ for
pairwise joint measurability).

In quantum theory, where ${\cal J}_{ij}$ is described by a joint POVM as defined above, the average anticorrelation probability $R_3$ takes the form~\cite{KG14}
\begin{equation}
R^Q_3=\frac{1}{3}\sum_{(ij)\in\{(12),(23),(13)\}}\text{Tr}\left[\left(G_{+-}^{ij}+G_{-+}^{ij}\right)\ket{\phi_0}\bra{\phi_0}\right],
\end{equation}
where $\ket{\phi_0}$ is the qubit state being measured. It follows that a necessary condition for state-dependent violation of the LSW inequality is
Tr$\left[\sum_{ij}(\alpha_{ij}\one-\vec{\sigma}\cdot\vec{a}_{ij})\ket{\phi_0}\bra{\phi_0}\right]<2\eta$.
It has been shown~\cite{KG14} that the largest violation of the LSW inequality for observables defined by \erf{trine}
can be obtained by the state
$\ket{\phi_0}=(\ket{0}+i\ket{1})/\sqrt{2}$,
and joint POVM ${\cal J}_{ij}$ in \erf{Gij} defined by
$\alpha_{ij}=1+\eta^2\hat{n}_i\cdot\hat{n}_j$ and a vector $\vec{a}_{ij}$ satisfying $\vec{a}_{ij}=(0,\sqrt{1+\eta^4(\hat{n}_i\cdot\hat{n}_j)^2-2\eta^2},0)$. Moreover, the optimal violation for $\eta$ in the range $[2/3,0.732]$ is as $\eta \to 2/3$,
so that $\alpha_{ij}\rightarrow 7/9$ and $|\vec{a}_{ij}|\rightarrow \sqrt{13}/9$ for any $(ij)\in \{(12),(23),(13)\}$.
Then the quantum average probability of anticorrelation is $R^Q_3\rightarrow 0.8114$ and exceeds the LSW noncontextual bound of $7/9\approx0.7778$.

In our experiment, we aim for $\eta=0.670$, strictly within the range $[2/3,0.732]$ but close to the optimum at $2/3$.

\begin{center}
\begin{figure*}
   \includegraphics[width=\textwidth]{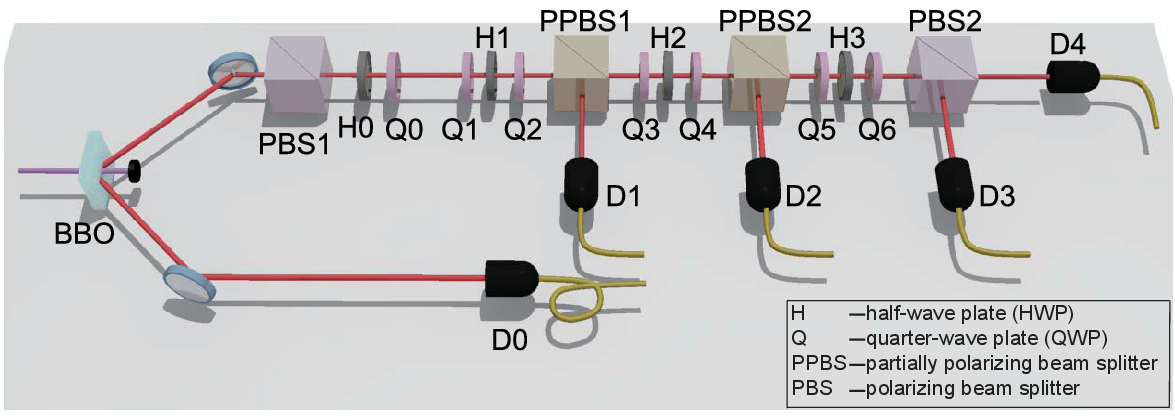}
   \caption{Experimental setup. Single photons are created via type-I spontaneous parametric down-conversion (SPDC) in
a $0.5$mm-thick nonlinear-$\beta$-barium-borate (BBO) crystal which is pumped by a CW diode laser with $80$mW of power. One photon in the pair is detected to herald the other photon, which is injected into the optical network. To realize the joint POVM ${\cal J}_{ij}$, the single-qubit rotations $U_{\epsilon\varepsilon}$ are realized by sandwich-type QWP-HWP-QWP sets with different angles placed in different optical modes. The first two projectors are realized by PPBSs with different transmission probabilities for vertical polarization. The last projector is realized by a PBS. Detecting heralded single photons means in practice registering coincidences between single photon detectors: D0 and each of D1-D4.}
 \label{setup}
\end{figure*}
\end{center}

\subsection{Implementation of joint POVMs}
For the pairwise joint measurements described above, each element of the POVM is rank one, and can be rewritten as $G^{ij}_{\epsilon\varepsilon}=\lambda_{\epsilon\varepsilon} \ket{\xi^{ij}_{\epsilon\varepsilon}}\bra{\xi^{ij}_{\epsilon\varepsilon}}$  with $\epsilon, \varepsilon\in\{+1,-1\}$. Here, $\lambda_{++}=\lambda_{--}=(2-\eta^2)/4$ and $\lambda_{+-}=\lambda_{-+} =(2+\eta^2)/4$.
We propose a scheme for implementing the joint POVMs in three stages, each of which is a single-qubit rotation followed by a two-outcome measurement. In each, the positive result (i.e.~a detector click) has a POVM element proportional to the appropriate projector, while the null result qubit is fed into the next stage. The null result qubit from the third stage is then also detected.

To be more specific, the three single-qubit rotations are designed as
\begin{equation}
U_{\epsilon\varepsilon}=\ket{0}\bra{\phi'_{\epsilon\varepsilon}}+\ket{1}\bra{\phi'^\perp_{\epsilon\varepsilon}},
\epsilon\varepsilon\in\{+-,-+,++\},
\end{equation}
while the POVM elements $\{P^0_{\epsilon\varepsilon},P^1_{\epsilon\varepsilon}\}$ take the form
\begin{equation}
P^0_{\epsilon\varepsilon}=\ket{0}\bra{0}+(1-\chi_{\epsilon\varepsilon})\ket{1}\bra{1},
P^1_{\epsilon\varepsilon}=\chi_{\epsilon\varepsilon}\ket{1}\bra{1}.
\end{equation}
In the above, $\ket{\phi'_{\epsilon\varepsilon}}$ and $\chi_{\epsilon\varepsilon}$ are chosen such that after the projector is applied, the probability of the click of $P^1_{\epsilon\varepsilon}$ is $\lambda_{\epsilon\varepsilon}\text{Tr}(\ket{\xi^{ij}_{\epsilon\varepsilon}}\bra{\xi^{ij}_{\epsilon\varepsilon}} \phi_0\rangle\bra{\phi_0})=\text{Tr}(G^{ij}_{\epsilon\varepsilon}\ket{\phi_0}\bra{\phi_0})$. If the other (null) result is obtained, the qubit enters the next stage of the apparatus, having had the operator $\sqrt{P^0_{\epsilon\varepsilon}}$ applied to its state. For example, we implement $G^{ij}_{+-}$ at the first stage by choosing $\ket{\phi'_{+-}}=|{\xi^{ij\perp}_{+-}}\rangle$ and $\chi_{+-}=\lambda_{+-}$. The first detector clicks with probability $\text{Tr}(G^{ij}_{+-}\ket{\phi_0}\bra{\phi_0})$, and if it does not click then the qubit state entering the next stage of
apparatus is $\sqrt{P^0_{+-}}U_{+-}\ket{\phi_0}=\langle\xi_{+-}^{ij\perp}\ket{\phi_0}\ket{0}+\sqrt{1-\chi_{+-}}\langle\xi^{ij}_{+-}\ket{\phi_0}\ket{1}$.
We design the apparatus so as to next measure $G^{ij}_{-+}$, then $G^{ij}_{++}$, in the same way.
 Since $\sum_{\epsilon\varepsilon}G^{ij}_{\epsilon\varepsilon}=\one$, the fourth possible click, following the null outcome at the third state, corresponds to the implementation of $G^{ij}_{--}$.
For further details see the Supplemental Material.

\begin{table*}[htbp]
%\scriptsize
\caption{Experimental results. The `element' columns contain the elements of joint POVM corresponding to heralded single-click events. The `probability' columns contain the measured probabilities of joint POVM. The `condition' column contains assigned
measurement values. The `value' column contains the measured probabilities of the anticorrelations. Rows 1-3 correspond to the results of joint POVMs on the state $\ket{\phi_0}$. Row 4 presents the measured $\tilde{R}^Q_3$. Error bar indicates the statistical uncertainty which is obtained based on assuming Poissonian statistics.}
\begin{tabular}{cccccccccc}
\hline
\multicolumn{2}{c}{D1} & \multicolumn{2}{c}{D2} & \multicolumn{2}{c}{D3} & \multicolumn{2}{c}{D4} & \multicolumn{2}{c}{Calculated contribution}\\
\cline{1-2}\cline{3-4}\cline{5-6}\cline{7-8}
element & probability & element & probability & element & probability & element & probability & condition & value \\
\hline
$G^{12}_{+-}$ & 0.4042(22) & $G^{12}_{-+}$ & 0.4065(22) & $G^{12}_{++}$ & 0.0948(11) & $G^{12}_{--}$ & 0.0944(10) & $\text{Pr}(X_1\neq X_2|G_{12})$ & 0.8108(15)\\

$G^{23}_{+-}$ & 0.4048(22) & $G^{23}_{-+}$ & 0.4067(22) & $G^{23}_{++}$ & 0.0953(11) & $G^{23}_{--}$ & 0.0931(10) & $\text{Pr}(X_2\neq X_3|G_{23})$ & 0.8116(15)\\

$G^{13}_{+-}$ & 0.4073(22) & $G^{13}_{-+}$ & 0.4078(22) & $G^{13}_{++}$ & 0.0931(11) & $G^{13}_{--}$ & 0.0919(10) & $\text{Pr}(X_1\neq X_3|G_{13})$ & 0.8150(15)\\
\hline
- & - & - & - & - & - & - & - & $\tilde{R}_3^{Q}$ & 0.8125(10)\\ \hline
\end{tabular}
\end{table*}

\section{Experimental Realization}

\subsection{Experimental violation of the LSW inequality} \label{sec:results}
We perform the test of the LSW inequality with single photons. The basis states of the qubit, $\ket{0}$ and $\ket{1}$, are encoded by the polarizations of single photons, $\ket{H}$ and $\ket{V}$. We generate contextual quantum correlations by performing the four-outcome joint POVM on this qubit.

The experimental setup shown in Fig.~\ref{setup} involves preparing the specific state (preparation stage) and then performing the joint POVM (measurement stage). In the preparation stage, polarization-degenerate trigger-herald photon pairs are produced and are registered by a coincidence count at two single-photon avalanche photodiodes (APDs) with $7$ns time window. Total coincidence counts are about $10^5$ over a collection time of $60$s, and the probability of randomly creating more than one simultaneous photon pair is thus of order $10^{-4}$, which is negligible. The second-order correlation $g^{(2)}$ is measured as $0.0089\pm0.0018$ which shows that the single-photon source is extremely non-classical~\cite{g2}. The heralded single photons are prepared in state $\ket{\phi_0}=(\ket{H}+i\ket{V})/\sqrt{2}$ after passing through a polarizing beam splitter (PBS), a half-wave plate (HWP, H0), and a quarter-wave plate (QWP, Q0).

In the measurement stage, to implement the two-outcome measurements, partially projecting polarizing elements are added to the setup and allow us to produce the required projectors with the appropriate weights. We employ partially polarizing beam splitters (PPBSs) with specific transmission probabilities for vertical polarization $T_{V}$ and same transmission probability for horizontal polarization $T_{H}=1$. This allows us to project the state onto $\ket{V}$ on the reflected port of the PPBSs.

After passing through each QWP-HWP-QWP set and the following PPBS, the photons are detected by APDs on the reflected port, in coincidence with the trigger photons. The transmitted photons go into the next QWP-HWP-QWP set and PPBS (or, at the final stage, a PBS which can be regarded as a special PPBS with equal transmission and reflection probabilities). The relative detection efficiencies of the detectors D2-D4 with respect to D1, are measured as $0.9499\pm0.0070$, $0.9199\pm0.0069$ and $0.9801\pm0.0063$ respectively and these figures are used to correct the coincidence counts (see the Supplemental Material). The probability of measuring the
photons is obtained by normalizing the corrected coincidence counts on each mode with respect to the total corrected coincidence counts. The overall detection efficiency of the heralded photons in our experiment is
approximately $11\%$. Thus we make the fair-sampling assumption: that the event selected out by the photonic coincidence is an unbiased representation of the whole sample.

The probabilities of photons being measured on the reflected ports (clicks on the detectors D1-D3) correspond to those of the joint POVM elements $G^{ij}_{+-}$, $G^{ij}_{-+}$, and $G^{ij}_{++}$, whereas the probability of photons being measured on the transmitted port of the PBS (click on the detector D4) corresponds to that of the element $G^{ij}_{--}$. We can estimate the matrix forms of the joint POVM elements from the measured probabilities (see Subsec.~\ref{sec:method}%Methods
). The negligible difference from the theoretical prediction guarantees successful experimental realization of joint POVMs by taking into account of all the imperfections of the experimental setup.

In Table I, we present the measured probabilities and the outcomes of the joint POVM with noise parameter $\eta=0.67$ on the specific state $\ket{\phi_0}$. The result of measured average probability of anticorrelations is $\tilde{R}^Q_3=0.8125\pm0.0010$. Here, and below, the tilde relates to the experimentally implemented POVMs, as opposed to the theoretical ones aimed for; see Subsec.~\ref{sec:method}. This $\tilde{R}^Q_3$  violates the bound set by the noncontextual hidden variable theory $1-\eta/3=0.7767$ by $35$ standard deviations. Furthermore, in our experiment the noise parameter can be estimated by the experimental data (see Subsec.~\ref{sec:method}). The average value of the estimated noise parameters in the experiment is $\overline{\tilde{\eta}}=0.6690\pm0.0019$. Using this value, rather than the aimed-for $0.670$, makes almost no difference in the LSW bound: the bound set by the noncontextual hidden variable theory can be calculated as $1-\overline{\tilde{\eta}}/3=0.7770\pm0.0006$ compared to $1-0.670/3=0.7767$. Even including the uncertainty in the former bound, the experimentally measured average probability of anticorrelation, $\tilde{R}^Q_3=0.8125\pm0.0010$,  still implies a violation of this experimental bound of the LSW inequality $1-\overline{\tilde{\eta}}/3$ by $22$ standard deviations. In Subsec.~\ref{sec:method} we give an alternate way of comparing the correlations and the bound, which also gives a violation by many standard deviations. Here, we finish by noting that the experimental value $\tilde{R}_3^Q$  is in agreement ($1.6$ standard deviations) with its theoretical prediction $0.8087\pm0.0022$, predicted via the estimated noise parameter $\overline{\tilde{\eta}}$.

\subsection{Evaluating the quality of experimental realization of POVM}\label{sec:method}
We consider the effect on the implementation of the joint POVM due to all the important imperfections, namely in the PPBSs ($T^1_V=0.3904\pm0.0045$, $T^2_V=0.2897\pm0.0050$), WPs (typical retardance accuracy$<2.67$nm), PBSs (typical extinction ratio $\sim10^5:1$), and detectors. We define a modified 2-norm distance $D(A,B)$ between the matrix form of the theoretical prediction of POVM element $A$ and that of experimental implementation of the corresponding POVM element $B$ as
\begin{equation}
D(A,B)=\sqrt{\frac{\text{Tr}\left[(A-B)^2\right]}{\text{Tr}(A^2)}}.
\label{eq:distance}
\end{equation}
For the particular forms of the POVM described in our paper, the distance ranges between $0$ for a perfect match and $\sqrt{2}$ for a complete mismatch. For example, we use the distance $D(G^{ij}_{\epsilon\varepsilon},\tilde{G}^{ij}_{\epsilon\varepsilon})$ to measure the mismatch between the theoretical prediction of $G^{ij}_{\epsilon\varepsilon}$ with $i\neq j\in\{1,2,3\}$, $\epsilon,\varepsilon\in\{+1,-1\}$, and the corresponding experimental implementation $\tilde{G}^{ij}_{\epsilon\varepsilon}$.

To obtain the distance, we perform  measurement tomography~\cite{F01,DPS13}. Single photons, prepared in the states $\ket{H}$, $\ket{V}$, $\ket{R}=(\ket{H}+i\ket{V})/\sqrt{2}$ and $\ket{D}=(\ket{H}+\ket{V})/\sqrt{2}$, are passed through the optical circuit and are detected by APDs in coincidence with the trigger photons. After correcting for the relative efficiencies of the different detectors, the photon counts give the measured probabilities. From these we can obtain the matrix forms of all twelve elements of the joint POVMs $\tilde{G}$ via maximum-likelihood estimation.

\begin{figure}
  \begin{center}
\includegraphics[width=.5\textwidth]{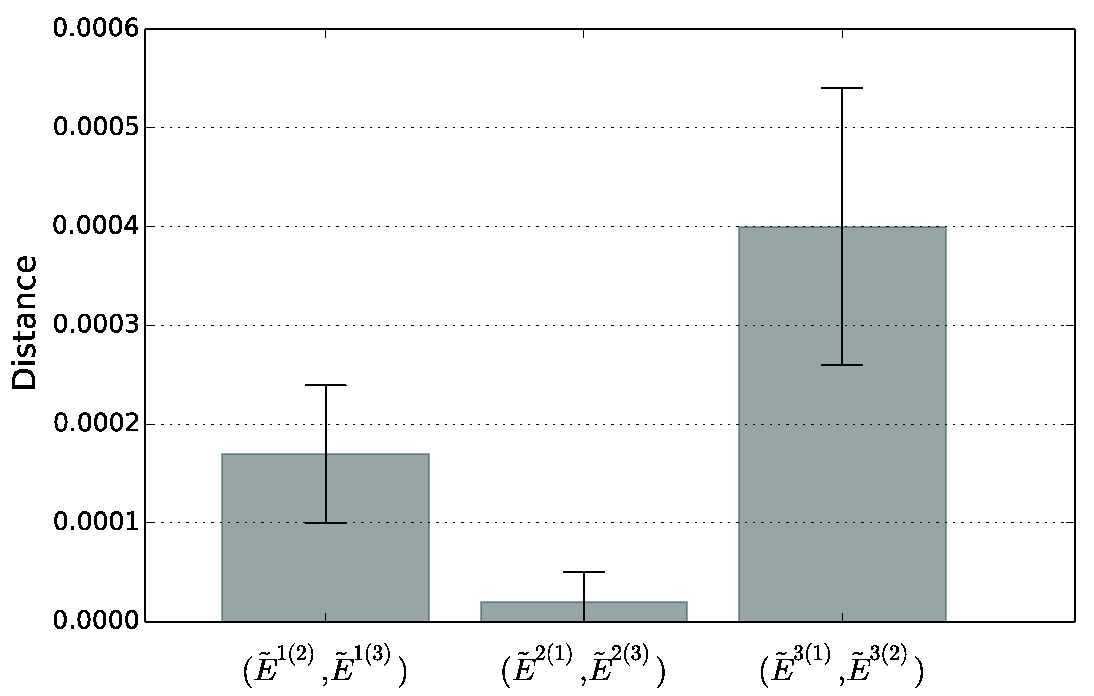}
\end{center}
\caption{The distance $D(\tilde{E}^{i(j)},\tilde{E}^{i(k)})$ between the estimated POVM elements for different contexts $\{{\cal M}_i,{\cal M}_j\}$ and $\{{\cal M}_i,{\cal M}_k\}$. Error bars indicate the statistical uncertainty, obtained from Monte-Carlo simulations assuming Poissonian photon-counting statistics.}
\label{fig:dist}
\end{figure}

In our experiment the accuracy of the experimental implementation of the measurements described by the POVM $\tilde{E}_\epsilon^{i(j)}$---the noisy version of the projective measurements---are more important. Here $\tilde{E}_{\epsilon}^{i(j)} \approx \sum_{\varepsilon}G_{\epsilon\varepsilon}^{ij}$, so the $(j)$ superscript indicates any dependence on context
in which it is performed (see below).
The element $\tilde{E}_{\epsilon}^{i(j)}$ is estimated by minimizing the 2-norm distance $D(\tilde{E}_{\epsilon}^{i(j)},\sum_{\varepsilon}\tilde{G}_{\epsilon\varepsilon}^{ij})$ which is defined in Eq.~(\ref{eq:distance}). This minimization is subject to the constraints that
the sum $\tilde{E}^{i(j)}_+ + \tilde{E}^{i(j)}_-$ equals the identity operator, whilst the difference is traceless, as per \erf{one-tr}. The minimum distances found by this procedure are very small (all less than $6.2\times10^{-5}$),
which justifies our approach. The theoretical prediction of the element satisfies the marginal condition $E^{i(j)}_{\epsilon}=\sum_{\varepsilon}G_{\epsilon\varepsilon}^{ij}=E^{i(k)}_\epsilon=\sum_{\varepsilon}G_{\epsilon\varepsilon}^{ik}$ ($i\neq j\neq k\in \{1,2,3\}$). However due to the imperfections in the experiment, there might be a slight difference between $\sum_{\varepsilon}\tilde{G}_{\epsilon\varepsilon}^{ij}$ and $\sum_{\varepsilon}\tilde{G}_{\epsilon\varepsilon}^{ik}$. Hence we use two superscripts $i$ and $j$ to represent the estimated element of POVM $\tilde{E}^{i(j)}_{\epsilon}$ which is estimated by $\tilde{G}^{ij}_{\epsilon\varepsilon}$ and corresponds to the joint measurable context $\{{\cal M}_i,{\cal M}_j\}$. The difference between the elements $\tilde{E}^{i(j)}_{\epsilon}$ and $\tilde{E}^{i(k)}_{\epsilon}$ can also be measured by the 2-norm distance $D(\tilde{E}^{i(j)}_{\epsilon},\tilde{E}^{i(k)}_{\epsilon})$ defined in Eq.~(\ref{eq:distance}).
Thus the distances satisfy $D(\tilde{E}^{i(j)}_+,\tilde{E}^{i(k)}_+)=D(\tilde{E}^{i(j)}_-,\tilde{E}^{i(k)}_-)$.  As shown in Fig.~\ref{fig:dist}, all the distances $D(\tilde{E}^{i(j)},\tilde{E}^{i(k)})$ are smaller than $0.0006$, which validates the experimental realizations of pairwise jointly measurable POVMs.

\begin{figure}[htbp]
\includegraphics[width=.5\textwidth]{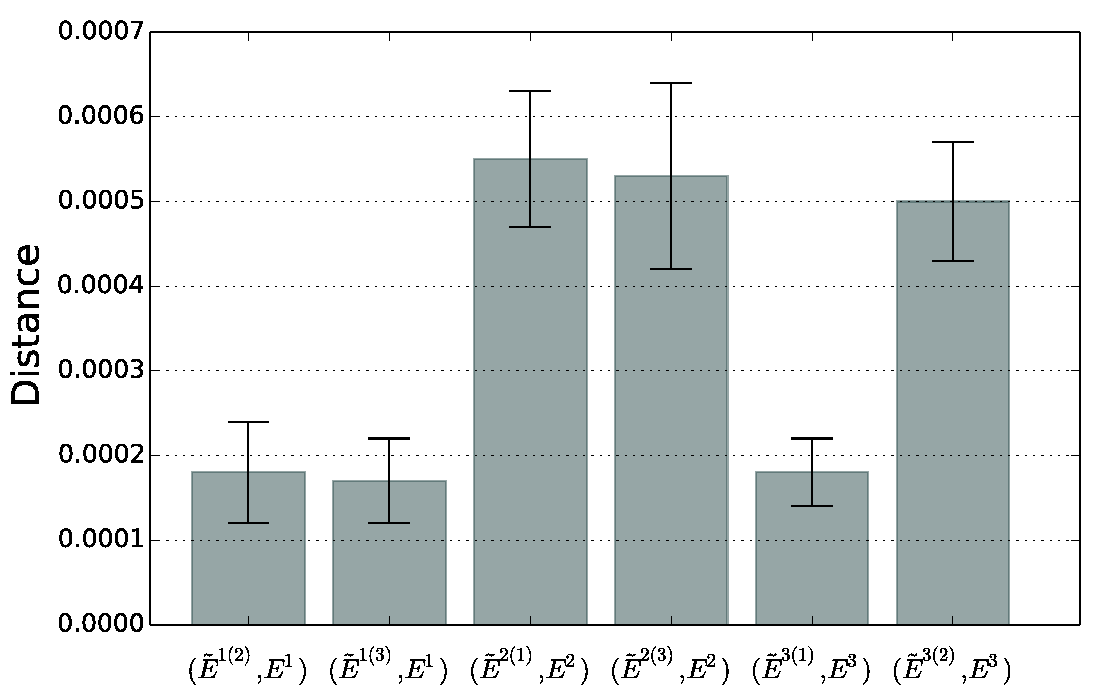}
\caption{The distance $D(E^{i},\tilde{E}^{i(j)})$ between the aimed-for POVM element $E^{i}$ for context $\{{\cal M}_i,{\cal M}_j\}$ and the estimated POVM elements from the experiment $\tilde{E}^{i(j)}$. Error bars indicate the statistical uncertainty, obtained from Monte-Carlo simulations assuming Poissonian photon-counting statistics.}
\label{fig:dist1}
\end{figure}

We also compare the estimated element $\tilde{E}^{i(j)}_\epsilon$ with the theoretical ideal $E^{i}_\epsilon$ by calculating the 2-norm distance $D(E^{i}_\epsilon,\tilde{E}^{i(j)}_\epsilon)$. Since the distances satisfy $D(E^{i}_+,\tilde{E}^{i(j)}_+)=D(E^{i}_-,\tilde{E}^{i(j)}_-)$ we only show six values of the distances $D(E^{i},\tilde{E}^{i(j)})$ in Fig.~\ref{fig:dist1}. All the distances are smaller than $0.0007$, which shows the successful experimental realizations of the POVMs with the chosen noise parameter $\eta=0.67$.

For determining the LSW bound used in Subsec.~\ref{sec:results} it is important to know the noise parameter of $\tilde{\eta}$ associated with the POVM. This can be estimated as
\begin{equation}
\tilde{\eta}^{i(j)}_\epsilon=\sqrt{2\text{Tr}\left[\tilde{E}^{i(j)}_\epsilon (\tilde{E}_\epsilon^{i(j)})^{\dagger}\right]-1}.
\end{equation}
The condition $\tilde{E}^{i(j)}_+ +\tilde{E}^{i(j)}_-=\mathds{1}$ guarantees that we have the same values of $\tilde{\eta}^{i(j)}_+$ and $\tilde{\eta}^{i(j)}_-$. Compared to the value of the noise parameter we aimed for in the experiment $\eta=0.67$, all the differences $|\tilde{\eta}^{i(j)}-\eta|$ are smaller than $0.0015$. The average value of the estimated noise parameters in the experiment is $\overline{\tilde{\eta}}=\frac{1}{6}\sum_{i(j)}\tilde{\eta}^{i(j)}=0.6690\pm0.0019$.

Finally, the value $R_3^Q$ corresponding to the ideal POVMs can also be bounded,  as follows. An arbitrary qubit POVM element $G$ can be written as $G=a \one+b  \hat{n} \cdot \vec{\sigma}$, where $\hat{n}$ is a unit vector and $a$ and $b$ are nonnegative numbers satisfying $b \leq a$ and $b \leq 1-a$. An arbitrary qubit density operator can be written as $\rho = \frac{1}{2} \left( \one+ \vec{r} \cdot \vec{\sigma} \right)$, where $|\vec{r}| \leq 1$. The probability of obtaining the outcome corresponding to $G$ for a POVM containing $G$ on state $\rho$ is given by $\text{Pr}(G) = \text{Tr} \left[ G \rho \right] = \frac{1}{2} \left( a + b \vec{r} \cdot \hat{n} \right)$.
Let $\vec{g} = (a,bn_x,bn_y,bn_z)$ and $\vec{s} = \frac{1}{2}(1,r_x,r_y,r_z)$. Then $\text{Pr}(G) = \vec{g} \cdot \vec{s}$. Let $\tilde{G}$ denote the experimentally realized POVM element corresponding to $G$, and likewise $\vec{\tilde{g}}$ to $\vec{g}$. Then $\left| \text{Pr}(G) - \text{Pr}(\tilde{G}) \right| =  \left| \left( \vec{g} - \vec{\tilde{g}} \right) \cdot \vec{s} \right| \leq  \left| \vec{g} - \vec{\tilde{g}} \right| \left| \vec{s} \right| \leq \frac{1}{\sqrt{2}} \left| \vec{g} - \vec{\tilde{g}} \right|$. Thus we obtain the bound
\begin{equation}\label{eq:boundPr}
\text{Pr}(G) \geq \text{Pr}(\tilde{G}) - \frac{1}{\sqrt{2}} \left| \vec{g} - \vec{\tilde{g}} \right| \;.
\end{equation}

Let $\tilde{g}_{+-}^{ij}$($\tilde{g}_{-+}^{ij}$) be the vector representation of $\tilde{G}_{+-}^{ij}$($\tilde{G}_{-+}^{ij}$) in our experiment, obtained above by tomography. Then from \eqref{eq:boundPr} we obtain a lower bound for the ideal value
\begin{align}
R_3^Q \geq &\tilde{R}_3^Q \notag\\
&- \frac{1}{3\sqrt{2}} \sum_{(ij)\in{(12),(23),(13)}} \left( \left| g_{+-}^{ij} - \tilde{g}_{+-}^{ij} \right| + \left| g_{-+}^{ij} - \tilde{g}_{-+}^{ij} \right| \right) \; . \label{corrterm}
\end{align}
We estimate the bound for the ideal value $R_3^Q$ based on the measured value $\tilde{R}_3^Q$ and the estimated $\tilde{G}_{\epsilon\varepsilon}^{ij}$. We find
\begin{align} \label{idealbound}
R_3^Q&\geq0.7964 \pm 0.0012.
\end{align}
The uncertainty here is larger than that in $\tilde{R}_3^Q$ because of uncertainties
in the $\tilde{g}$s that contribute to the correction term in \erf{corrterm}. Now, the appropriate
point of comparison is the ideal noncontextual bound of $0.7767$, from the aimed-for
$\eta = 0.67$, because we are inferring the correlations from
an ideal measurement with this $\eta$.
The value of the bound in \erf{idealbound} implies a violation of this
ideal bound  by at least
$16$ standard deviations.

Note that as we assume the validity of quantum mechanics, there is no need to establish operational equivalences between the measured POVM elements in different contexts, as done in Ref.~\cite{Mazurek:2016aa}.

\section{Discussion}
Any realistic measurement necessarily has some nonvanishing amount of noise and therefore never achieves the ideal of sharpness. This provides a compelling reason to test contextuality applicable to unsharp measurements. Here we test the generalized noncontextuality inequality for the unsharp measurements of LSW~\cite{LSW}.
For unsharp measurements that can be jointly performed,
correlated noise could allow correlations to be generated by a noncontextual hidden
variable model. The LSW inequality takes such
correlations into account by setting a higher bound. Thus a violation of the LSW inequality certifies nonclassicality that cannot be attributed to hidden variables associated with noise in the unsharp measurements.

Our experimental results show convincing violation of the LSW inequality with single-photon qubits. That is, it is a demonstration of contextuality for the simplest type of quantum system. It is also the first experiment to apply the LSW argument to rule out noncontextuality within quantum theory.

The experimental confirmation of quantum contextuality in its simple and fundamental form sheds new light on the contradiction between quantum mechanics and noncontextual realistic models. Furthermore, we realize joint POVMs of noisy spin-$\frac{1}{2}$ observables on a single-qubit system which is the key point to implement the unsharp measurements, paving the way for further developments such as the real time estimation~\cite{CG09}, monitoring of the Rabi oscillations of a single qubit in a driving field~\cite{KU12} and understanding the relation between information gain and disturbance~\cite{W11}.

\section*{Funding Information}
NSFC (Nos. 11474049, 11674056 and 11674306), NSFJS (No. BK20160024), Scientific Research Foundation of Graduate School of Southeast University (No. YBJJ1623), Australian Research Council Centre of Excellence CE110001027 and grant (FQXi-RFP-610 1504) from the Foundational Questions Institute Fund (fqxi.org) at the Silicon Valley Community Foundation.

\section*{Acknowledgments}

We would like to thank R. Kunjwal for stimulating discussions, and three anonymous reviewers for helpful suggestions.

\section*{Supplemental Documents}
See Supplement 1 for supporting content.

% Bibliography
\bibliography{optica_qubit_contextuality}

\clearpage

\setcounter{figure}{0}
\makeatletter
\renewcommand{\thefigure}{S\@arabic\c@figure}
\makeatother

\setcounter{equation}{0}
\makeatletter
\renewcommand{\theequation}{S\@arabic\c@equation}
\makeatother

\setcounter{table}{0}
\makeatletter
\renewcommand{\thetable}{S\@arabic\c@table}
\makeatother
\setcounter{page}{1}
\setcounter{section}{0}

\pagebreak

\begin{widetext}
\begin{center}

\textbf{\large Experimental generalized contextuality with single-photon qubits: supplementary material}

\vspace{3mm}
In the supplementary document, we provide the details of the experiment and data analysis. The raw probabilities to demonstrate the joint measurability of positive operator-valued measures (POVMs) are also provided.

\vspace{3mm}

\end{center}

\section{Implementation of the elements of the joint POVM $G$}

Each element of the POVM can be written as $G^{ij}_{\epsilon\varepsilon}=\lambda_{\epsilon\varepsilon}\ket{\xi^{ij}_{\epsilon\varepsilon}}\bra{\xi^{ij}_{\epsilon\varepsilon}}$, with $\epsilon,\varepsilon\in\{+1,-1\}$, $i\neq j\in\{1,2,3\}$,
\begin{equation}\label{eq:1}
\lambda_{++}=\lambda_{--}=\frac{(2-\eta^2)}{4}, \lambda_{+-}=\lambda_{-+}=\frac{(2+\eta^2)}{4},
\end{equation} and
\begin{equation}
 \ket{\xi^{12}_{+-}}=\left(
 \begin{array}{c}
   \sqrt{3}\eta+i\sqrt{4-8\eta^2+\eta^4}\\
   -2+3\eta-\eta^2
 \end{array}\right)/\sqrt{8-12\eta+8\eta^2-6\eta^3+2\eta^4},\notag\end{equation}\begin{equation}
   \ket{\xi^{12}_{-+}}=\left(
   \begin{array}{c}
     \sqrt{3}\eta-i\sqrt{4-8\eta^2+\eta^4}\\
     2+3\eta+\eta^2
     \end{array}\right)/\sqrt{8+12\eta+8\eta^2+6\eta^3+2\eta^4},\notag\end{equation}\begin{equation}
     \ket{\xi^{12}_{++}}=\left(
     \begin{array}{c}
       \sqrt{3}\eta+i\sqrt{4-8\eta^2+\eta^4}\\
       2-\eta-\eta^2
       \end{array}\right)/\sqrt{8-4\eta-8\eta^2+2\eta^3+2\eta^4},\notag\end{equation}\begin{equation}
       \ket{\xi^{12}_{--}}=\left(
       \begin{array}{c}
         \sqrt{3}\eta-i\sqrt{4-8\eta^2+\eta^4}\\
         -2-\eta+\eta^2
         \end{array}\right)/\sqrt{8+4\eta-8\eta^2-2\eta^3+2\eta^4},\notag\end{equation}\begin{equation}
  \ket{\xi^{13}_{+-}}=\left(
         \begin{array}{c}
           \sqrt{3}\eta-i\sqrt{4-8\eta^2+\eta^4}\\
           2-3\eta+\eta^2
           \end{array}\right)/\sqrt{8-12\eta+8\eta^2-6\eta^3+2\eta^4},\notag\end{equation}\begin{equation}
    \ket{\xi^{13}_{-+}}=\left(
           \begin{array}{c}
             -\sqrt{3}\eta+i\sqrt{4-8\eta^2+\eta^4}\\
             2+3\eta+\eta^2
             \end{array}\right)/\sqrt{8+12\eta+8\eta^2+6\eta^3+2\eta^4},\notag\end{equation}\begin{equation}
    \ket{\xi^{13}_{++}}=\left(
             \begin{array}{c}
               \sqrt{3}\eta-i\sqrt{4-8\eta^2+\eta^4}\\
               -2+\eta+\eta^2
               \end{array}\right)/\sqrt{8-4\eta-8\eta^2+2\eta^3+2\eta^4},\notag\end{equation}\begin{equation}
    \ket{\xi^{13}_{--}}=\left(
               \begin{array}{c}
                 -\sqrt{3}\eta+i\sqrt{4-8\eta^2+\eta^4}\\
                 -2-\eta+\eta^2
                 \end{array}\right)/\sqrt{8+4\eta-8\eta^2-2\eta^3+2\eta^4},\notag\end{equation}\begin{equation}
    \ket{\xi^{23}_{+-}}=\left(
                 \begin{array}{c}
                   \sqrt{12}\eta-i\sqrt{4-8\eta^2+\eta^4}\\
                   2+\eta^2
                   \end{array}\right)/\sqrt{2}(2+\eta^2),\notag\end{equation}\begin{equation}
        \ket{\xi^{23}_{-+}}=\left(
                   \begin{array}{c}
                     -\sqrt{12}\eta-i\sqrt{4-8\eta^2+\eta^4}\\
                     2+\eta^2
                     \end{array}\right)/\sqrt{2}(2+\eta^2),\notag\end{equation}\begin{equation}
        \ket{\xi^{23}_{++}}=\left(
                     \begin{array}{c}
                       -i\sqrt{4-4\eta-4\eta^2+2\eta^3+\eta^4}\\
                       -\sqrt{4+4\eta-4\eta^2-2\eta^3+\eta^4}
                       \end{array}\right)/\sqrt{2}(-2+\eta^2),\notag\end{equation}\begin{equation}
            \ket{\xi^{23}_{--}}=\left(
                       \begin{array}{c}
                         -i\sqrt{4+4\eta-4\eta^2-2\eta^3+\eta^4}\\
                         -\sqrt{4-4\eta-4\eta^2+2\eta^3+\eta^4}
                         \end{array}\right)/\sqrt{2}(-2+\eta^2),\notag
\end{equation}
Each element can be implemented by a single-qubit rotation followed by a two-outcome measurement. In the first step, to realize the element $G_{+-}^{ij}$ the single-qubit rotation is designed as
\begin{equation}\label{eq:2}
  U_{+-}=\ket{0}\bra{\phi_{+-}^\prime}+\ket{1}\bra{\phi_{+-}^{\prime\perp}}, \ket{\phi_{+-}^\prime}=\ket{\xi_{+-}^{ij\perp}}.
\end{equation}
The POVM element $\{P_{+-}^0,P_{+-}^1\}$ takes the form
\begin{equation}\label{eq:3}
  P_{+-}^0=\ket{0}\bra{0}+(1-\chi_{+-})\ket{1}\bra{1},P_{+-}^1=\chi_{+-}\ket{1}\bra{1}, \chi_{+-}=\lambda_{+-}.
\end{equation}
The choice of $\ket{\phi_{+-}^\prime}$ guarantees if $P^1_{+-}$ clicks the initial state is projected onto the eigenstate $\ket{\xi_{+-}^{ij}}$ and let the component of the state $\ket{\xi^{ij\perp}_{+-}}$ which is orthogonal to $\ket{\xi_{+-}^{ij}}$ all pass through for the next measurement. Therefore the state after the first rotation is $U_{+-}\ket{\phi_0}\bra{\phi_0}U_{+-}^\dagger$ for any input $\ket{\phi_0}$. The first detector ($P^1_{+-}$) clicks  with the probability $\text{Tr}(P_{+-}^1 U_{+-}\ket{\phi_0}\bra{\phi_0}U_{+-}^\dagger)=\lambda_{+-}\bra{\xi_{+-}^{ij}}\phi_0\rangle\langle\phi_0\ket{\xi_{+-}^{ ij}}=\text{Tr}(G_{+-}^{ij}\ket{\phi_0}\bra{\phi_0})$. Thus we implement the element of POVM $G_{+-}^{ij}$. The other state without click is
\begin{equation}\label{eq:4}
\sqrt{P_{+-}^0}U_{+-}\ket{\phi_0}=\innertime{\xi_{+-}^{ij\perp}}{\phi_0}\ket{0}+\sqrt{1-\chi_{+-}}\innertime{\xi_{+-}^{ij}}{\phi_0}\ket{1}.
\end{equation}

In the second step, to implement the element $G_{-+}^{ij}$, the single-qubit rotation is designed as
\begin{equation}\label{eq:5}
  U_{-+}=\ket{0}\bra{\phi_{-+}^\prime}+\ket{1}\bra{\phi_{-+}^{\prime\perp}},
\end{equation}
where
\begin{equation}\label{eq:6}
\ket{\phi_{-+}^\prime}=\frac{1}{\mathcal{N}_{-+}}\left(\innertime{\xi_{+-}^{ij\perp}}{\xi_{-+}^{ij\perp}}\ket{0}+\sqrt{1-\chi_{+-}}\innertime{\xi_{+-}^{ij}}{\xi_{-+}^{ij\perp}}\ket{1}\right)
\end{equation}
with the normalization factor $\mathcal{N}_{-+}$. Compared Eqs.~(\ref{eq:5}) and (\ref{eq:7}), one can see that we only replace $\ket{\phi_0}$ in Eq.~(\ref{eq:5}) by $\ket{\xi_{-+}^{ij\perp}}$ in Eq.~(\ref{eq:7}). The second POVM element takes the form $\{P_{-+}^0,P_{-+}^1\}$, where
\begin{equation}\label{eq:7}
  P_{-+}^0=\ket{0}\bra{0}+(1-\chi_{-+})\ket{1}\bra{1},P_{-+}^1=\chi_{-+}\ket{1}\bra{1}, \chi_{-+}=\frac{\lambda_{-+}\mathcal{N}_{-+}^2}{1-\lambda_{+-}}.
\end{equation}
The parameter $\chi_{-+}$ is chosen such that after the projector is performed, the probability of the click of $P_{-+}^1$ is that of the measurement $G_{-+}^{ij}$ performing on the initial state $\ket{\phi_0}$. The state after the second qubit rotation is $U_{-+}P_{+-}^0U_{+-}\ket{\phi_0}\bra{\phi_0}U_{+-}^\dagger U_{-+}^\dagger$. The second detector ($P^1_{-+}$) clicks with the probability $\text{Tr}(P_{-+}^1 U_{-+}P_{+-}^0U_{+-}\ket{\phi_0}\bra{\phi_0}U_{+-}^\dagger U_{-+}^\dagger)=\lambda_{-+}\bra{\xi_{-+}^{ij}}\phi_0\rangle\langle\phi_0\ket{\xi_{+-}^{ij}}=\text{Tr}(G_{-+}\ket{\phi_0}\bra{\phi_0})$. Thus we implement the element $G_{-+}^{ij}$.

The other state without click is
\begin{align}\label{eq:8}
&\sqrt{P_{-+}^0}U_{-+}\sqrt{P_{+-}^0}U_{+-}\ket{\phi_0}\\
&=\left(\innertime{\phi_{-+}^\prime}{0}\innertime{\xi_{+-}^{ij\perp}}{\phi_0}+\sqrt{1-\chi_{+-}}\innertime{\phi_{-+}^{\prime}}{1}\innertime{\xi_{+-}^{ij}}{\phi_0}\right)\ket{0} \nonumber \\
&+\sqrt{1-\chi_{-+}}\left(\innertime{\phi_{-+}^{\prime\perp}}{0}\innertime{\xi_{+-}^{ij\perp}}{\phi_0}+\sqrt{1-\chi_{+-}}\innertime{\phi_{-+}^{\prime\perp}}{1}\innertime{\xi_{+-}^{ij}}{\phi_0}\right)\ket{1}.\nonumber
\end{align}

In the third step, to implement $G_{++}^{ij}$ the single-qubit rotation is designed as
 \begin{equation}\label{eq:9}
   U_{++}=\ket{0}\bra{\phi_{++}^\prime}+\ket{1}\bra{\phi_{++}^{\prime\perp}},
 \end{equation}
where
\begin{align}\label{eq:10}
\ket{\phi_{++}^\prime}=&\frac{1}{\mathcal{N}_{++}}\Big[\big(\innertime{\phi_{-+}^\prime}{0}\innertime{\xi_{+-}^{ij\perp}}{\xi_{++}^{ij\perp}}+\sqrt{1-\chi_{+-}}\innertime{\phi_{-+}^{\prime}}{1}\innertime{\xi_{+-}^{ij}}{\xi_{++}^{ij\perp}}\big)\ket{0}\\
\nonumber&+\sqrt{1-\chi_{-+}}\big(\innertime{\phi_{-+}^{\prime\perp}}{0}\innertime{\xi_{+-}^{ij\perp}}{\xi_{++}^{ij\perp}}+\sqrt{1-\chi_{+-}}\innertime{\phi_{-+}^{\prime\perp}}{1}\innertime{\xi_{+-}^{ij}}{\xi_{++}^{ij\perp}}\big)\ket{1}\Big]
\end{align}
with the normalization factor $\mathcal{N}_{++}$. Comparing Eqs.~(\ref{eq:9}) and (\ref{eq:11}), one can find that we replace $\ket{\phi_0}$ in Eq.~(\ref{eq:9}) by $\ket{\xi^{ij\perp}_{++}}$ in Eq.~(\ref{eq:11}). The third POVM element takes the form $\{P_{++}^0,P_{++}^1\}$, where
 \begin{equation}\label{eq:11}
   P_{++}^0=\ket{0}\bra{0},P_{++}^1=\ket{1}\bra{1}.
 \end{equation}
 With this setup when the input state is $\ket{\xi_{++}^{ij\perp}}$ the detector ($P_{++}^1$) never clicks. Thus the measurement corresponding to click of $P_{++}^1$ is proportional to $\ket{\xi_{++}^{ij}}\bra{\xi_{++}^{ij}}$. We now prove the measurement we implement is exactly $G_{++}^{ij}$.

Assume the POVM elements corresponding to clicks of $P_{++}^1$ and $P_{++}^0$ are $x\ket{\xi_{++}^{ij}}\bra{\xi_{++}^{ij}}$ and $y\ket{\psi}\bra{\psi}$. Due to the fact that we have realized $\lambda_{+-}\ket{\xi_{+-}^{ij}}\bra{\xi_{+-}^{ij}}$ and $\lambda_{-+}\ket{\xi_{-+}^{ij}}\bra{\xi_{-+}^{ij}}$, we have
 \begin{equation}
      \begin{aligned}
        x\ket{\xi_{++}^{ij}}\bra{\xi_{++}^{ij}}+y\ket{\psi}\bra{\psi}&=\mathds{1}-\lambda_{+-}\ket{\xi_{+-}^{ij}}\bra{\xi_{+-}^{ij}}-\lambda_{-+}\ket{\xi_{-+}^{ij}}\bra{\xi_{-+}^{ij}}\\
        &=\lambda_{++}\ket{\xi_{++}^{ij}}\bra{\xi_{++}^{ij}}-\lambda_{--}\ket{\xi_{--}^{ij}}\bra{\xi_{--}^{ij}}
      \end{aligned}
      \label{eq:12}
\end{equation}

Tracing both sides of Eq.~(\ref{eq:12}) leads to
\begin{equation}\label{eq:13}
  x+y=\lambda_{++}+\lambda_{--},
\end{equation}
and
\begin{equation}\label{eq:14}
 \bra{\xi_{++}^{ij\perp}}\Big(x\ket{\xi_{++}^{ij}}\bra{\xi_{++}^{ij}}+y\ket{\psi}\bra{\psi}\Big)\ket{\xi_{++}^{ij\perp}}=\bra{\xi_{++}^{ij\perp}}\Big(\lambda_{++}\ket{\xi_{++}^{ij}}\bra{\xi_{++}^{ij}}-\lambda_{--}\ket{\xi_{--}^{ij}}\bra{\xi_{--}^{ij}}\Big)\ket{\xi_{++}^{ij\perp}}
\end{equation} leads to
\begin{equation}\label{eq:15}
  y=\lambda_{--}
\end{equation}
Then we also have
\begin{equation}\label{eq:16}
  x=\lambda_{++}
\end{equation}
Thus the POVM element corresponding to the click of $P_{++}^1$ is $G_{++}^{ij}=\lambda_{++}\ket{\xi_{++}^{ij}}\bra{\xi_{++}^{ij}}$. The POVM element corresponding to the click of $P_{++}^0$ is $G_{--}^{ij}=\lambda_{--}\ket{\xi_{--}^{ij}}\bra{\xi_{--}^{ij}}$.

\section{the measurement stage of realizing joint POVMs}

In the measurement stage, the single-qubit rotations can be realized by a combination of quarter-wave plates (QWPs) and half-wave plates (HWPs), so-called a sandwich-type QWP-HWP-QWP set, with certain setting angles depending on the parameters of the joint positive operator-valued measure (POVM). The setting angles of the wave plates (WPs) used to realize the corresponding elements of joint POVMs are shown in Table~\ref{tab:1}. We employ partially polarizing beam splitters (PPBSs) with specific transmission probabilities for vertical polarization $T_{V}$ and same transmission probability for horizontal polarization $T_{H}=1$. This allows us to implement the measurement $\bra{V}$ on the reflected port of the PPBSs.

In the basis $\{\ket{H},\ket{V}\}$, the single-qubit rotations realized by HWP and QWP are
\begin{align}\label{eq:17}
&R_{HWP}(\theta_H)=\begin{pmatrix}\cos2\theta_H & \sin2\theta_H\\
\sin2\theta_H & -\cos2\theta_H\end{pmatrix},\\\nonumber
&R_{QWP}(\theta_Q)=\begin{pmatrix}\cos^2\theta_Q+i \sin^2 \theta_Q & (1-i)\sin\theta_{Q}\cos\theta_Q\\
(1-i)\sin\theta_Q\cos\theta_Q & \sin^2\theta_Q+i\cos^2\theta_Q \end{pmatrix},
\end{align}
respectively, where $\theta_H$ and $\theta_Q$ are the angles between the optic axes of HWP and QWP and horizontal direction.

\begin{table}[htbp]%[tbp]
\small
\caption{The setting angles of WPs and the transmission probabilities for vertical polarization $T^{1(2)}_V$ of the two PPBSs for realization of joint POVMs.}
\begin{center}
  \begin{tabular}{ccccccccc}
  \hline
  $G_{ij}$ & $\theta_{Q1}/\theta_{Q2}$ & $\theta_{H1}$ & $\theta_{Q3}/\theta_{Q4}$ & $\theta_{H2}$ & $\theta_{Q5}/\theta_{Q6}$ & $\theta_{H3}$ & $T^1_V$ & $T^2_V$ \\
  \hline
  $G_{12}$ & $-7.3^\circ$ & $32.9^\circ$ & $1.4^\circ$ & $-29.5^\circ$  & $22.4^\circ$ & $69.2^\circ$ & $0.3904(45)$ & $ 0.2897(50)$\\
  $G_{23}$ & $21.7^\circ$ & $60.2^\circ$ & $-4.9^\circ$ & $-36.5^\circ$ & $21.7^\circ$ & $60.2^\circ$ & $0.3904(45)$  & $ 0.2897(50)$ \\
  $G_{13}$ & $7.3^\circ$ & $47.5^\circ$ & $-1.4^\circ$ & $-32.2^\circ$ & $-112.4^\circ$ & $110.8^\circ$ & $0.3904(45)$  & $ 0.2897(50)$\\ \hline
  \end{tabular}

\end{center}
\label{tab:1}
\end{table}

\section{relative detection efficiencies of detectors}

After applying joint POVM on the single-photon qubit, the photons are detected by single-photon avalanche photodiodes (APDs) on the reflected ports of the two PPBSs, and both reflected and transmitted ports of PBS, in coincidence with the trigger photons. The relative detection efficiencies of the detectors D1-D4 are measured and used to correct the coincidence counts.

To measure the relative efficiencies of the different detectors D1, D2, D3 and D4, we make a reasonable assumption that the total number of photons is fixed. We tune the setting angles of WPs to change the photon distribution. That is, for each time after we tune the setting angles of WPs, the normalized number of photons at each output port with respect to the total number of photons is changed, which can be read at the corresponding detector. The readout photon counts at each detector equal to the number of photons at each output port multiplied by the relative efficiency of the corresponding detector. After we tune the setting angles of WPs for four times, we have four linear equations with four variables (relative efficiencies of detectors) and then solve them to obtain the relative efficiencies of the detectors D1, D2, D3, and D4.

In our experiment, the relative efficiencies of the detectors D1, D2, D3, and D4 are $1$, $0.9499\pm0.0070$, $0.9199\pm0.0069$ and $0.9801\pm0.0063$ respectively calculated from the experimental data. Thus we can use the relative efficiencies of the detectors to correct the photon counts in the measurement stage. For example, after the correction the photon counts at D2 which are used to calculate the probability of the photons being measured at D2 should be the readout photon counts divided by the relative efficiency of D2 $0.9499\pm0.0070$.

\section{Joint measurability of POVMs.}

We test the joint measurability of the constructed joint POVM $G_{ij}$. For different qubit states, we analyze the experimental results of the joint POVM and test whether the marginal condition $\sum_{X_{j(i)}} G_{X_i,X_j}^{ij}=E_{X_{i(j)}}^{i(j)}$ is satisfied. Without loss of generality, we choose the four states $|H\rangle$, $|V\rangle$, $|R\rangle=(|H\rangle+i|V\rangle)/\sqrt{2}$, and $|D\rangle=(|H\rangle+|V\rangle)/\sqrt{2}$ as states being measured. The results are shown in Table~\ref{tab:2}.

The measured probabilities Tr$(\sum_{X_j}G^{ij}_{X_i,X_j}\rho)$ are in agreement with the theoretical predictions of the probabilities Tr$(E_{X_i}^i\rho)$ of the POVM element $E_{X_i}^i$ on the state $\rho\in\{\ket{H}\bra{H},\ket{V}\bra{V},\ket{R}\bra{R},\ket{D}\bra{D}\}$, which proves the marginal condition is satisfied. Thus the constructed joint POVM $G_{ij}$ shows the joint measurability.

\newpage

\begin{table}[htbp]
%\scriptsize
\caption{Experimental results of $\sum_{X_j}G^{ij}_{X_i,X_j}$ on the four different single-qubit states compared to the theoretical predictions of the probabilities of $E_{X_i}^i$ on these states.}
\begin{center}
  \begin{tabular}{ccccc}
  \hline
  \quad & $|H\rangle$ & $|V\rangle$ & $|R\rangle$ & $|D\rangle$\\
  \hline
  $G_{++}^{12}+G_{+-}^{12}$ & 0.8357(13) & 0.1607(13) & 0.4990(21) & 0.4958(20)\\
  $G_{++}^{13}+G_{+-}^{13}$ & 0.8353(13) & 0.1699(13) & 0.5003(22) & 0.5029(21)\\
  $E_+^1\text{(Theory)}$ & 0.8350 & 0.1650 & 0.5000 & 0.5000\\
  \hline
  $G_{-+}^{12}+G_{--}^{12}$ & 0.1642(13) & 0.8393(13) & 0.5009(21) & 0.5042(20)\\
  $G_{-+}^{13}+G_{--}^{13}$ & 0.1647(13) & 0.8301(13) & 0.4997(22) & 0.4971(21)\\
  $E_-^1\text{(Theory)}$ & 0.1650 & 0.8350 & 0.5000 & 0.5000\\
  \hline
  $G_{++}^{12}+G_{-+}^{12}$ & 0.3155(20) & 0.6747(19) & 0.5013(22) & 0.7818(17)\\
  $G_{++}^{23}+G_{+-}^{23}$ & 0.3179(19) & 0.6748(19) & 0.5001(21) & 0.7788(16)\\
  $E_+^2\text{(Theory)}$ & 0.3325 & 0.6675 & 0.5000 & 0.7901\\
  \hline
  $G_{+-}^{12}+G_{--}^{12}$ & 0.6845(20) & 0.3253(19) & 0.4987(22) & 0.2182(17)\\
  $G_{-+}^{23}+G_{--}^{23}$ & 0.6821(19) & 0.3252(19) & 0.4999(21) & 0.2212(16)\\
  $E_-^2\text{(Theory)}$ & 0.6675 & 0.3325 & 0.5000 & 0.2099\\
  \hline
  $G_{++}^{13}+G_{-+}^{13}$ & 0.3454(21) & 0.6652(19) & 0.5008(22) & 0.2083(16)\\
  $G_{++}^{23}+G_{-+}^{23}$ & 0.3448(20) & 0.6519(21) & 0.5020(22) & 0.2102(17)\\
  $E_+^3\text{(Theory)}$ & 0.3325 & 0.6675 & 0.5000 & 0.2099\\
  \hline
  $G_{+-}^{13}+G_{--}^{13}$ & 0.6546(21) & 0.3348(19) & 0.4992(22) & 0.7917(16)\\
  $G_{+-}^{23}+G_{--}^{23}$ & 0.6551(20) & 0.3481(21) & 0.4980(22) & 0.7898(17)\\
  $E_-^3\text{(Theory)}$ & 0.6675 & 0.3325 & 0.5000 & 0.7901\\
  \hline
  \end{tabular}

\end{center}
\label{tab:2}
\end{table}

\end{widetext}
\end{document}